\documentclass[aps,pre,
 onecolumn,
 12pt,
 notitlepage,
 showpacs,floatfix,nofootinbib,
 double-space
 superscriptaddress
 ]{revtex4-2}

 \usepackage{subcaption}
 \usepackage{bbold}
 \usepackage{amssymb}
 \usepackage{amsfonts}
 \usepackage{amsmath}
 \usepackage{amsthm}
 \usepackage{epsfig}
 \usepackage{float}
 \usepackage{multirow}
 \usepackage[usenames,dvipsnames]{color}
 \usepackage[latin1]{inputenc}
 \usepackage{xr}
 \usepackage{enumitem}
 \usepackage{microtype}
 \usepackage{graphics,graphicx,xcolor}
 \usepackage{comment}
 \usepackage[hidelinks,unicode=true]{hyperref}
 \usepackage[normalem]{ulem}
 \usepackage{lipsum}
 \usepackage{cancel}
\usepackage[hidelinks,unicode=true]{hyperref}
\hypersetup{colorlinks=true,
	linkcolor=blue,
	urlcolor=blue,
	citecolor=blue,
	pdfhighlight=/N
}

\newcommand{\sig}[2]{\vec{\sigma}_{#1}^{#2}}
\newcommand{\kk}[1]{\mathbf{K}_{#1}}
\newcommand{\gum}[0]{\mathcal{G}_1}
\newcommand{\gdo}[0]{\mathcal{G}_2}

\graphicspath{{newfigs/}}

\begin{document}

\title{Dynamics of matrix coupled Kuramoto oscillators on modular networks: excitable behavior and global decoherence}

\author{Guilherme S. Costa}
\affiliation{ICTP South American Institute for Fundamental Research \& Instituto de Física Teórica - UNESP, 01140-070, São Paulo, Brazil }

\author{Marcus A. M. de Aguiar}
\affiliation{ICTP South American Institute for Fundamental Research \& Instituto de Física Teórica - UNESP, 01140-070, São Paulo, Brazil }
\affiliation{Instituto de Física "Gleb Wataghin", Universidade Estadual de Campinas, Unicamp 13083-970, Campinas, SP, Brazil}

\begin{abstract}
	Synchronization is observed in many natural systems, with examples ranging from neuronal activation to walking pedestrians. The models proposed by Winfree and Kuramoto stand as the classic frameworks for investigating these phenomena. The Kuramoto model, in particular, has been extended in different ways since its original formulation to account for more general scenarios. One such extension replaces the coupling parameter with a coupling matrix, describing a form of generalized frustration with broken rotational symmetry. A key feature of this model is the existence of {\it phase tuned states}, characterized by having the phase of the order parameter  pointing in the direction of the dominant eigenvector of the coupling matrix. Here we investigate the matrix coupled Kuramoto model on networks with two modules, such that one module is in the phase tuned state and the other in a state where the order parameter rotates. We identified different regimes in which one or the other module dominates the dynamics. We found, in particular, that the phase tuned module can create a bottleneck for the oscillation of the rotating module, leading to a behavior similar to the charge and fire regimes of excitable systems. We also found an extended region in the parameter space where motion is globally disordered, even though one of the modules presented high levels of synchronization when uncoupled.
\end{abstract}
	
\maketitle




\section{INTRODUCTION}

Networks with modular structure are found in many branches of the social sciences and biology. From contact communities in social networks to functional areas of the brain \cite{porter2009,Winding2023,Sanchez2021}, the way nodes are connected affect the underlying dynamics of the system. Spreading of diseases and rumors are examples of well studied systems that are strongly affected by the network topology \cite{Cota2018,Maia2021,moreno2002epidemic,de2018fundamentals}. Moreover, many of these systems are governed by periodic processes exhibiting different levels of synchronization, such as pacemakers cells in the heart, fireflies and neurons in the brain. The dynamics of synchronization have also been shown to depend on the heterogeneity and modular organization of the nodes \cite{Oh2005,Park2006,Skardal2012,Ujjwal2016}.

In the case of neuronal networks, simple models for the oscillatory behavior of neurons are provided by the coupled phase oscillator theories developed by Winfree \cite{Winfree1967} and  Kuramoto \cite{kuramoto1975}. One key interest in the study of these systems is to understand if synchronization in modular networks can happen independently in each module and under what conditions it occurs throughout  the network leading to global synchronization \cite{Gomez2011,Bick2020,Pecora2014}. Epilepsy, for instance, has been related to the synchronous fire of large portions of the brain \cite{jiruska2013synchronization}.  Furthermore, recent studies show that information processing in the cerebral cortex is associated with synchronization mechanisms \cite{Tognoli2014} while some brain disorders such as Alzheimer have been associated with abnormal neural sync \cite{Uhlhaas2006}.

In Winfree's model, the oscillators are described only by their phases $\theta_i$  and are coupled according to \cite{Winfree1967,Strogatz2000} 
\begin{equation}
	\dot{\theta_i} = \omega_i + Q(\theta_i) \sum_{j=1}^{N}P(\theta_j).
\end{equation}  
In numerical simulations, the natural frequencies $\omega_i$ are  usually extracted from a symmetric distribution $g(\omega)$; $P(\theta_j)$ is the {\it pulse function} produced by oscillator $j$ and $Q(\theta_i)$ is the response of oscillator $i$ to the pulse field generated by the $N$ oscillators. This model was extensively studied and applied to several problems with different functional forms of $P$ and $Q$ \cite{Pazo2014,Okeeffe2016,Chandra2017,Gallego2017}.

Kuramoto showed later that, for weak coupling and nearly identical oscillators, the equation for the phases could also be written as  \cite{kuramoto1975} 
\begin{equation}
	\label{eq::kuramotoOriginal}
	\dot{\theta_i} = \omega_i + \dfrac{K}{N} \sum_{j=1}^{N} \operatorname{sin}(\theta_j - \theta_i),
\end{equation}
where $K$ is the coupling strength. The sinusoidal coupling is actually the simplest case of a more general class of coupling functions involving the phase differences $(\theta_j-\theta_i)$ \cite{Strogatz2000}. Several extensions of the Kuramoto model have been proposed since its original formulation, such as oscillators embedded on networks of different topologies \cite{Gomez2011,francisco2016}, subjected to external forces \cite{childs2008,moreira2019}, coupled with particle motion \cite{OKeeffe2017,Lizarraga2023} and higher dimensional phase spaces \cite{chandra2019,zhu2013}.

In both Winfree and Kuramoto models, the degree of synchronization can be measured by the complex order parameter
\begin{equation}
	\label{eq::defz}
	z = pe^{i\psi} = \sum_{j=1}^{N}e^{i\theta_j},
\end{equation}
in which $p \approx 1 $ implies phase synchronization whereas $p \approx 0$ indicates decoherent motion. If each oscillator is represented as a point on the unit circle at position $\theta_i$, then the vector $\vec{p}= (p\cos\psi,p\sin\psi)$ represents the center of mass of the particles. In the limit of $N \to \infty$ it can be shown \cite{kuramoto1975} that the Kuramoto model exhibits a continuous phase transition from disorder to synchronization as $K$ increases. For $K < K_c = 2/\pi g(0)$, the oscillators move as if they were independent ($p = 0$) and for $K > K_c$ they start to cluster and $p$ increases as $p = \sqrt{1 - K_c/K}$. 

In this paper we consider an extension of the Kuramoto model where the oscillators are described by unit vectors $\vec{\sigma}_i=(\cos\theta_i,\sin\theta_i)$ \cite{chandra2019} and the coupling constant is promoted to a coupling matrix acting on these vectors \cite{Buzanello2022}. This extension will be discussed in detail in Section II, where we show that it  reduces to versions of Winfree, Kuramoto and Kuramoto-Sakaguchi models depending on how the coupling matrix is chosen. 

The model was previously studied in the context of homogeneous interactions \cite{Buzanello2022} and higher dimensions \cite{Aguiar2023,Fariello2024}. One key feature of this formulation is the existence of {\it phase tuned states}, where $\vec{p}$, the center of mass, converges to the direction of the dominant eigenvector of the coupling matrix. Here we take advantage of this property and investigate the model in networks with two modules, setting one of the modules in a phase tuned state and the other in a state of synchronized rotation. We show that when the modules are connected the system displays a rich and complex dynamic behavior depending on the inter-module coupling strength. In particular, we found that the phase tuned module slows down the rotation of the second module as its center of mass passes through the phase tuned direction, and speeds it up in the perpendicular direction, leading to a behavior similar to that found in excitable systems. We also found a region in the parameter space where the motion is globally disordered, even though the uncoupled dynamics in one of the modules presented high levels of synchronization. 

The paper is organized as follows: in Section  \ref{sec::modMatrixCoup}, we review the matrix coupled Kuramoto model and discuss its connection with previous synchronization models. We  derive the dynamical equations for the case of modular interactions by performing a dimensional reduction using the Ott-Antonsen ansatz in Section \ref{sec::modModGraph}. Constraining our analysis to systems with two modules, we investigate the phase diagrams for some particular types of coupling in Section \ref{sec::distinctMod}. In Section \ref{sec::simulations} we perform simulations on synthetic modular systems for comparative purposes. Finally, some discussions and remarks are presented in Section \ref{sec::discussion}.

\section{MATRIX COUPLING}
\label{sec::modMatrixCoup}

In this section we extend the Kuramoto model following the ideas introduced in \cite{chandra2019,Buzanello2022}. The first step is
to describe the oscillators  by unit vectors $\vec{\sigma_i} = (\cos \theta_i,\sin \theta_i)$. The vectors satisfy the equations 
\begin{equation}
	\label{eq::vecform}
	\dfrac{d \sig{i}{}}{dt} = \textbf{W}_i \sig{i}{} + \dfrac{K}{N} \sum_{j=1}^{N} [ \sig{j}{} - (\sig{i}{}\cdot \sig{j}{}) \sig{i}{}],
\end{equation}
where 
\begin{equation}
	\textbf{W}_i = 
	\begin{pmatrix} 
		0 & \omega_i \\
		-\omega_i &0 
	\end{pmatrix}.
\end{equation}

Eqs. (\ref{eq::vecform}) are completely equivalent to Eqs.(\ref{eq::kuramotoOriginal}). The advantage of writing the equations in this format 
is that $\vec{\sigma_i}$ can be naturally extended to $D$-dimensional unit vectors, described by $D-1$ spherical angles,
instead of a single phase $\theta_i$. The only requirement is that $ \textbf{W}_i$ are $D$-dimensional anti-symmetric matrices \cite{chandra2019}. 
Here we shall keep $D=2$ but will promote the coupling constant $K$ to a matrix $\kk{}$ acting on the vectors as \cite{Buzanello2022}
\begin{equation}
	\dfrac{d \sig{i}{}}{dt} = \textbf{W}_i \sig{i}{} + \dfrac{1}{N} \sum_{j=1}^{N} [ \mathbf{K}\sig{j}{} - (\sig{i}{}\cdot \mathbf{K} \sig{j}{}) \sig{i}{}].
	\label{eq::kuraFrust}
\end{equation}

The coupling matrix can be interpreted as a form of generalized frustration, in which the vector $\sig{j}{}$ is rotated by $\kk{}$ before interacting with $\sig{i}{}$.  In this formulation, the order parameter can be calculated as 
\begin{equation}
	\vec{p} = \dfrac{1}{N}\sum_{i=1}^{N} \vec{\sigma}_i = (p\cos \psi,p\sin \psi).
	\label{eq::orderP}
\end{equation}

It is useful to write $\mathbf{K}$  as a sum of a rotation matrix $\kk{R}$ and a symmetric matrix $\kk{S}$ as
\begin{equation}
	\textbf{K} \equiv \kk{R} + \kk{S} =  K 
	\begin{pmatrix}
		\cos \alpha       & \sin \alpha  \\
		-\sin \alpha & \cos \alpha 
	\end{pmatrix}
	+ J
	\begin{pmatrix}
		-\cos \beta       & \sin \beta  \\
		\sin \beta & \cos \beta 
	\end{pmatrix}.
	\label{matrizK}
\end{equation}
The corresponding equations for the phases $\theta_i$ are 
\begin{equation}
	\dot{\theta}_i = \omega_i + \sum_{j=1}^{N} \left[ K \sin (\theta_i - \theta_j - \alpha) + J \sin(\theta_i + \theta_j + \beta)\right].
	\label{eq::thetai}
\end{equation}

For $J \neq 0$, the system exhibits novel behaviors such as active states, where the phase and module of $\vec{p}$ oscillate in time,  and phase tuned states, in which all oscillators lock onto the direction of the dominant eigenvector of $\kk{}$, breaking the rotational symmetry of the system. 

Interestingly, this formulation encompasses several previous models of coupled oscillators. For $J = \alpha = 0$, we recover the original Kuramoto model while setting only $J = 0$ the equations reduce to the Kuramoto-Sakaguchi version \cite{Sakaguchi1986}. Finally, for $J = K$, the model reduces to a version of the Winfree equations with $Q(\theta) = -2\sin{(\theta + \alpha_-)}$ and $P(\theta) = \cos{(\theta + \alpha_+)}$, where $\alpha_- = (\beta -\alpha)/2$ and $\alpha_+ = (\beta +\alpha)/2$, similar to the problem investigated in \cite{Manoranjani2023} for $\alpha = \beta = 0$. 

Using Eq.(\ref{eq::orderP}) and defining $\vec{q} \equiv \kk{} \vec{p} \equiv  q (\cos\gamma, \sin \gamma)$, Eq.(\ref{eq::kuraFrust}) can also be written as
\begin{equation}
	\dfrac{d \sig{i}{}}{dt} = \textbf{W}_i \sig{i}{} +  [ \vec{q} - (\sig{i}{}\cdot \vec{q}) \sig{i}{}]
	\label{eq::kuraFrust2}
\end{equation}
and Eq.(\ref{eq::thetai}) for the phases becomes
\begin{equation}
		\dot{\theta}_i = \omega_i  +  q \sin(\gamma - \theta_i).
\end{equation}
Notice that for $J=\alpha=0$, $\kk{}$ is proportional to the identity, leading to $q = K p$ and $\gamma = \psi$.

\section{MODULAR GRAPHS}
\label{sec::modModGraph}

\subsection{General equations}

To investigate the interactions between oscillators in modular structures, we consider a network divided into $\Omega$ communities
and write Eq. \eqref{eq::kuraFrust} as
\begin{equation}
	\label{eq::dynamic}
	\frac{d \sig{i}{s}}{dt} = \textbf{W}_i \sig{i}{s} + \sum_{m=1}^{\Omega} \dfrac{\eta_m}{N_{m}} \sum_{j=1}^{N_{m}} [ \mathbf{K}_{sm}\sig{j}{m} - (\sig{i}{s}\cdot \mathbf{K}_{sm} \sig{j}{m}) \sig{i}{s}],
\end{equation}
where oscillator $i$ belongs to module $s = 1,2,...,\Omega$. Here $\eta_m = N_m/N$ is the fraction of oscillators within module $m$, and the matrices $\kk{sm}$ are parametrized by $K_{sm}$, $J_{sm}$, $\alpha_{sm}$ and $\beta_{sm}$ as in Eq. \eqref{matrizK}. In addition, we define order parameters for each module as
\begin{equation}
	\vec{p}_s = \left( p_s \cos{\psi_s}, p_s \sin{\psi_s} \right) = \dfrac{1}{N_s}\sum_{i=1}^{N_s}\sig{i}{s},
\end{equation}
and for the whole system as
\begin{equation}
	\vec{p}_T = \left( p_T \cos{\psi_T}, p_T \sin{\psi_T} \right) = \sum_{s=1}^{\Omega}\eta_s \vec{p}_{s}.
\end{equation}

As in the case of a single module, we define the auxiliary vectors
\begin{equation}  
	\label{eq::defq}
\vec{q}_{sm} = \left( q_{sm} \cos{\gamma_{sm}}, q_{sm} \sin{\gamma_{sm}} \right) = \dfrac{\eta_s}{N_{s}} \sum_{j=1}^{N_{s}} \mathbf{K}_{sm}\sig{j}{s} =  \eta_s \mathbf{K}_{sm}\vec{p}_s,
\end{equation}
to rewrite Eq \eqref{eq::dynamic} as
\begin{equation}
	\label{eq::dynamic2}
	\frac{d \sig{i}{s}}{dt} = \textbf{W}_i \sig{i}{s} + \sum_{m=1}^{\Omega} [\vec{q}_{sm} - (\sig{i}{s}\cdot \vec{q}_{sm}) \sig{i}{s}]
\end{equation}
or, in terms of the phases,
\begin{equation}
	\label{eq::thetaim}
	\dot{\theta}_{is} = \omega_{is} + \sum_{m=1}^{\Omega} q_{sm} \sin(\gamma_{sm} - \theta_{is}).
\end{equation}

\subsection{Ott-Antonsen ansatz}
	
Next, we consider the limit of infinite oscillators and  use the Ott-Antonsen ansatz \cite{Ott2008} to write differential equations for the order parameters $\vec{p}_s$ instead of individual oscillators. We define $f_s(\omega,\theta,t)$ as the density of oscillators belonging to module $s$ with natural frequency $\omega$ at position $\theta$ in time $t$. The density satisfies the continuity equation
\begin{equation}
	\label{eq::continuity}
	\dfrac{\partial f_s}{\partial t} + \dfrac{\partial (f_s v_{s \theta })}{\partial \theta} = 0,
\end{equation}
with velocity field
\begin{equation}
	\label{eq::velocity}
	v_{s \theta } = \omega_s + \sum_{m=1}^{\Omega} q_{sm} \sin(\gamma_{sm} - \theta) = \omega_s + \dfrac{1}{2i} \left( He^{-i\theta} - H^*e^{i\theta} \right),
\end{equation}
where
\begin{equation}
H = \sum_{m=1}^{\Omega}q_{sm} e^{i\gamma_{sm}}.
\end{equation}

The ansatz consists in expanding $f_s$ in Fourier series and  choosing the coefficients in terms of a single complex parameter $\nu_s(\omega,t)$:
\begin{equation}
	\label{eq::density}
	f_s(\omega,\theta,t) = \dfrac{g_s(\omega)}{2\pi} \left[ 1 + \sum_{n=1}^{\infty}\nu_s^n e^{-i n \theta} + c.c. \right ].
\end{equation} 

Writing $z_s = p_s e^{i\psi_s}$ and taking the continuum limit we obtain
\begin{equation}
	\label{eq::zspolo}
	z_s = \int f_s(\omega,\theta,t) e^{i\theta}d\theta d\omega = \int g_s(\omega) \nu_s(\omega) d\omega,
\end{equation}
where we used (\ref{eq::density}). This equation can be solved analytically if $g_s(\omega)$ is a Lorentzian distribution:
\begin{equation}
	g_s(\omega) = \dfrac{1}{\pi}\dfrac{\Delta_s}{(\omega - \omega_{0s})^2 + \Delta_s^2}.
\end{equation}

The integrand has poles at $\omega = \omega_{0s} \pm i\Delta_s$ and the overall integral can be performed by using the residues theorem, resulting in  $z_s = \nu_s(\omega_{0s}+i\Delta_s)$ \cite{Ott2008}.

Returning to the continuity equation we use Eqs. \eqref{eq::velocity} and \eqref{eq::density} to calculate 
\begin{align*}
	\dfrac{\partial f_s}{\partial t} &= \dfrac{g_s(\omega)}{2\pi} \left[ \sum_{n=1}^{\infty} n \dot{\nu}_s \nu_s^{n-1}e^{-in\theta} + c.c. \right], \\
	\dfrac{\partial f_s}{\partial \theta} &= -\dfrac{g_s(\omega)}{2\pi} \left[ \sum_{n=1}^{\infty} in  \nu_s^{n}e^{-in\theta} + c.c. \right], \\
	\dfrac{\partial v_{s \theta }}{\partial \theta} &= -\dfrac{1}{2} \left( He^{-i\theta} + H^*e^{i\theta} \right).
\end{align*}

Substituting in Eq. \eqref{eq::continuity} we obtain a differential equation for the ansatz parameters $\nu_s$ 
\begin{equation}
	\label{eq::dyneta}
	\dot{\nu}_s =  i\omega_s \nu_s  + \dfrac{H}{2} - \dfrac{H^*}{2}\nu_s^2.
\end{equation}

Calculating $\omega_s$ at $(\omega_{0s}+i\Delta_s)$ we can replace $\nu_s$  by $z_s$ for Lorentz distributions. Also, 
using Eq. \eqref{eq::defq}, we obtain
\begin{equation}
	\label{eq::H}
H = \sum_{m=1}^{\Omega} K_{sm} \eta_m z_{m}e^{-i\alpha_{sm}} - J_{sm}\eta_m z_{m}^*e^{-i\beta_{sm}}
\end{equation}
resulting in 
\begin{equation}
	\label{eq:zcomplex2}
	\dot{z}_s = i(\omega_{s0} + i\Delta_s)z_s - \sum_{m=1}^{\Omega}\dfrac{\eta_m}{2} \left[  \left( K_{sm} z^*_{m}e^{i\alpha_{sm}} -  J_{sm}z_{m}e^{i\beta_{sm}}\right) z_s^2  + \left(  K_{sm} z_{m}e^{-i\alpha_{sm}} - J_{sm}z_{m}^*e^{-i\beta_{sm}} \right) \right].
\end{equation}

Finally, separating real and imaginary parts of Eq. \eqref{eq:zcomplex2}, we find the dynamical equations for the module and phase of the modular order parameter $z_s$ as
\begin{align}
	\dot{p}_s &= -\Delta_s p_s + \dfrac{(1-p_s^2)}{2}\sum_{m=1}^{\Omega}\eta_m p_m \left[ K_{sm}\cos (\psi_s - \psi_m + \alpha_{sm})  - J_{sm}\cos (\psi_s + \psi_m + \beta_{sm})\right], \\
	p_s\dot{\psi}_s &=  \omega_{0s} p_s - \dfrac{(1+p_s^2)}{2}\sum_{m=1}^{\Omega}\eta_m p_m \left[ K_{sm}\sin (\psi_s - \psi_m + \alpha_{sm})  + J_{sm}\sin (\psi_s + \psi_m + \beta_{sm})\right].
\end{align}

From now on, we will consider the particular case of two modules of the same size. In this case, the factors $\eta_m=1/2$ can be absorbed into $K_{sm}$ and $J_{sm}$. The most general coupling between two modules requires 4 matrices: $\kk{11}$, $\kk{22}$, $\kk{12}$ and $\kk{21}$, resulting in sixteen parameters. For sake of simplicity, we will consider that matrices $\kk{12}$ and $\kk{21}$ are identical and given by $K_{12}\mathbb{1}$.  In addition, we set $\beta = 0$ for all $\kk{}$ \cite{Buzanello2022}.  With these assumptions, the equations describing the system can be written as 
\begin{align}
	\dot{p}_1 &= -\Delta_1p_1 + \frac{p_1}{2}(1-p_1^2) \left [ K_{1}\operatorname{cos}\alpha_{1} - J_{1}\operatorname{cos}(2\psi_1) \right] + \frac{p_2}{2}(1-p_1^2)K_{12} \operatorname{cos}\xi \label{eq::ansatzP1}\\		
	\dot{p}_2 &= -\Delta_2p_2 + \frac{p_2}{2}(1-p_2^2) \left [ K_{2}\operatorname{cos}\alpha_{2} - J_{2}\operatorname{cos}(2\psi_2) \right] + \frac{p_1}{2}(1-p_2^2)K_{12} \operatorname{cos}\xi  \label{eq::ansatzP2}\\
	p_1\dot{\psi}_1 &= +\omega_1p_1 - \frac{p_1}{2}(1+p_1^2) \left [ K_{1}\operatorname{sin}\alpha_{1} - J_{1}\operatorname{sin}(2\psi_1) \right] - \frac{p_2}{2}(1+p_1^2)K_{12} \operatorname{sin} \xi 	\label{eq::ansatzPhi1} \\			
	p_2\dot{\psi}_2 &= +\omega_2p_2 - \frac{p_2}{2}(1+p_2^2) \left [ K_{2}\operatorname{sin}\alpha_{2} - J_{2}\operatorname{sin}(2\psi_2) \right] + \frac{p_1}{2}(1+p_2^2) K_{12} \operatorname{sin}\xi \label{eq::ansatzPhi2}		
\end{align}
in which $\xi = \psi_1 - \psi_2$ is the phase difference between the modules. 

\section{DYNAMICS OF TWO MODULES}
\label{sec::distinctMod}

Despite the reduction in the number of parameters, Eqs.  \eqref{eq::ansatzP1} to \eqref{eq::ansatzPhi2} are still complex and difficult to solve analytically for $J_i \neq 0$. In this paper we shall focus on the interactions between a module governed by the Kuramoto-Sakaguchi equations, where the local order parameter rotates, and one  in the phase tuned state, where the order parameter freezes in a specific direction. Moreover, we shall keep the intensity of the internal couplings within the modules fixed, and vary only the inter-module intensity $K_{12}$ and the frustration parameter $\alpha_1$. To reduce the number of parameters even more we also fix $\omega_1=\omega_2=0$ and $\Delta_1=\Delta_2=1$.

In order to assess the role of symmetry breaking parameter $J_2$ in the dynamics we consider the following situations:

\noindent {\it case 1} - $\gum$: $K_1=10$, $J_1=0$; \qquad $\gdo$: $K_2=0$, $J_2=3$, $\alpha_2=0$;

\noindent {\it case 2} - $\gum$: $K_1=10$, $J_1=0$; \qquad $\gdo$: $K_2=3$, $J_2=0$, $\alpha_2=0$. 

These choices correspond to modules displaying very distinct behaviors when isolated: in both cases the oscillators of module $\mathcal{G}_1$ synchronize with $\dot{\psi}_1 \approx K_1 \sin\alpha_1$. In the first case module $\gdo$ will be in phase tuned state \cite{Buzanello2022}, in which $\vec{p}_2$ aligns with the principal eigenvector of $\kk{2}$ (which is the $\hat{y}$ direction for $\beta=0$).  In the second case, on the other hand, the rotational symmetry is restored and $\psi_2$ converges to a random angle that depends on the initial conditions.  

With these considerations, we performed numerical integration of Eqs. \eqref{eq::ansatzP1} to \eqref{eq::ansatzPhi2} using fourth order Runge-Kutta algorithm, varying the phase-lag parameter $\alpha_1$ and the inter-module coupling $K_{12}$, computing the stationary time-average order parameter for both individual modules ($\langle p_1 \rangle$ and $\langle p_2 \rangle$) and the system as a whole ($\langle p_T \rangle$). 

\begin{figure}[ht]
	\centering
	\includegraphics[width=0.9\textwidth,trim={0 4.8cm 0 0},clip]{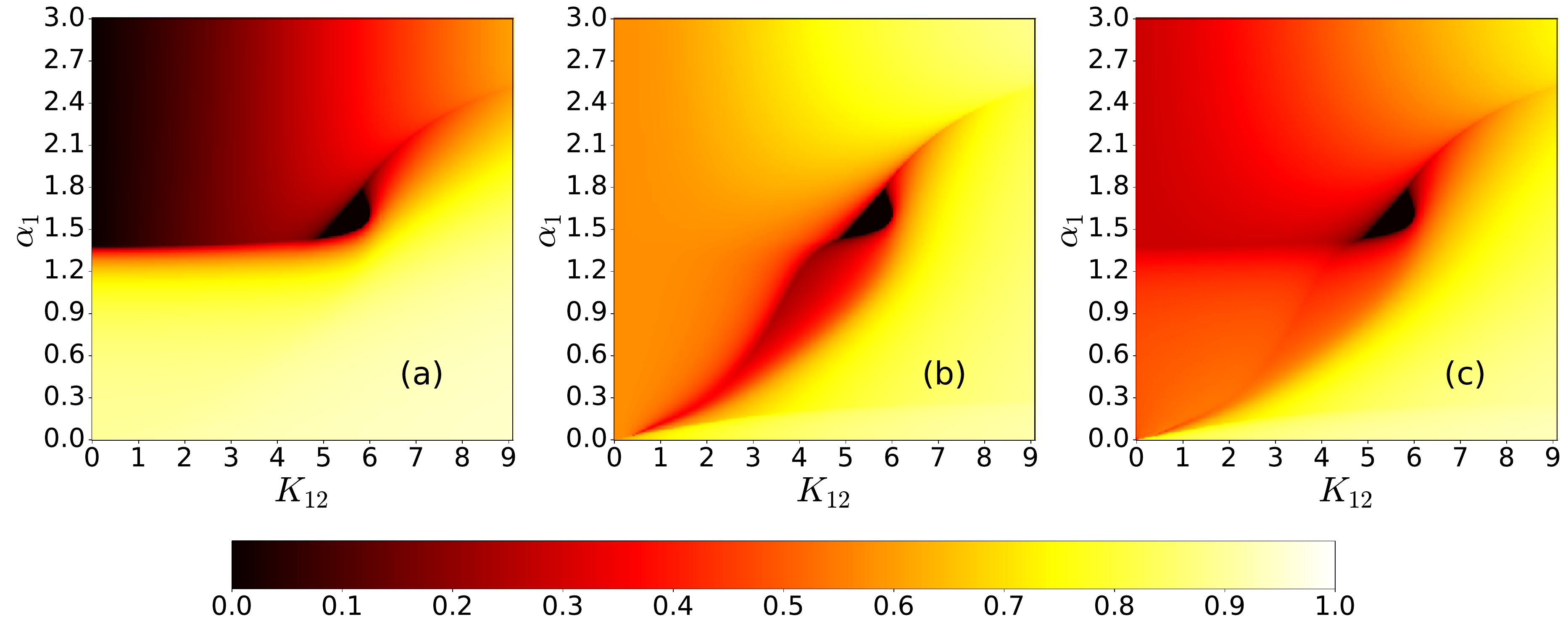}
	\includegraphics[width=0.9\textwidth,trim={0 2cm 0 0},clip]{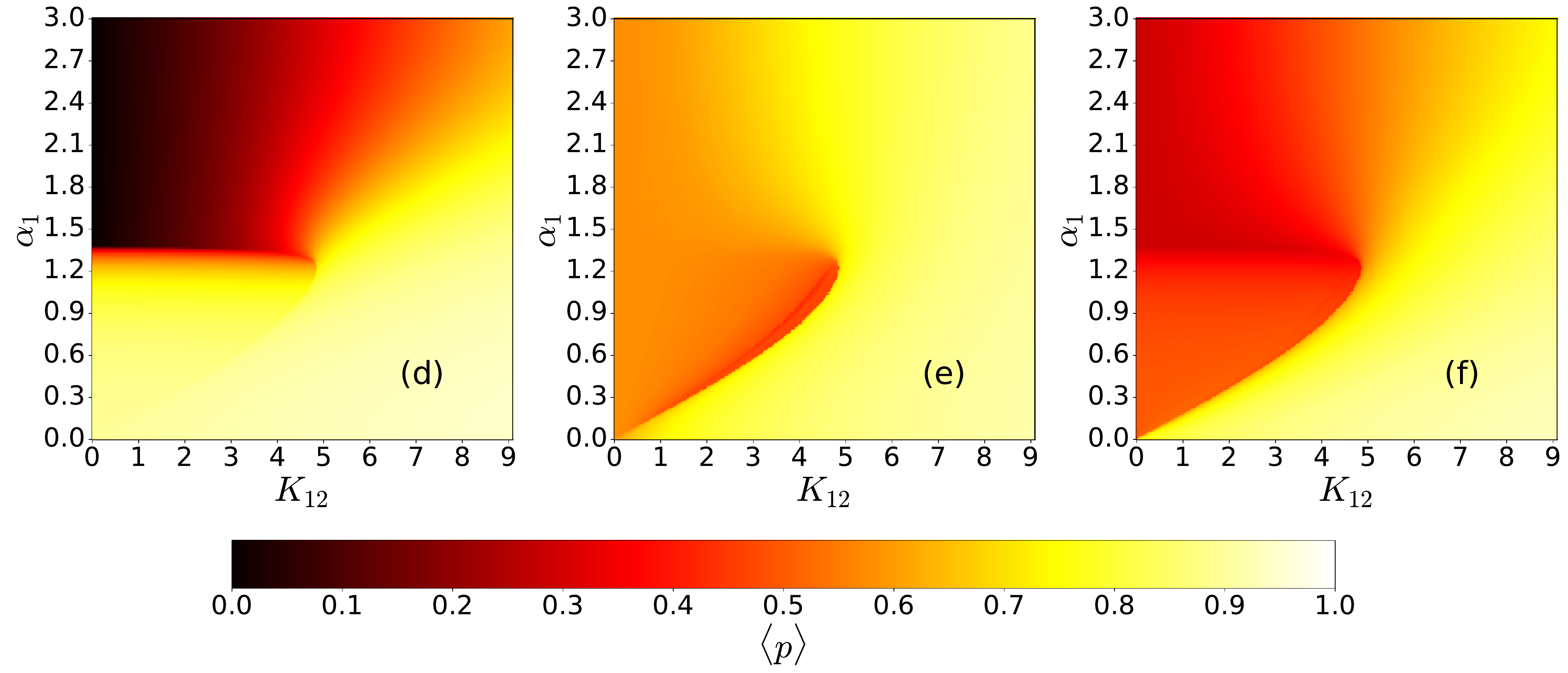}
	\caption{Heatmaps in $\alpha_1 \times K_{12}$ parameter space showing the time averaged order parameters $\langle p_1 \rangle$ (a and d); $\langle p_2 \rangle$ (b and e); and $\langle p_t \rangle$ (c and f).  Panels (a) to (c) refer to case 1 and (d) to (f) to case 2. }
\label{fig::ansatzK10J3}
\end{figure}

Figures \ref{fig::ansatzK10J3}(a)-(c) show the average orders parameters for $\gum$, $\gdo$ and the full system $\gum + \gdo$ in  the form of a heatmap in the $\alpha_1 \times K_{12}$ plane for case 1 whereas Figures \ref{fig::ansatzK10J3}(d) and (e) are the counterparts for case 2. Figures \ref{fig::ansatzK10J3}(a) and \ref{fig::ansatzK10J3}(d), that refer to $\gum$, present a transition for small $K_{12}$ at $\alpha_1 \approx 1.37$, in accordance with the usual Kuramoto-Sakaguchi order-disorder transition, in which the solution $p_1 = 0$ becomes stable when $K \cos \alpha < 2 \Delta$ \cite{Tanaka2014}. For larger values of $K_{12}$ this behavior changes, as the modular structure becomes fuzzy. 

\begin{figure}[ht]
	\centering
	\includegraphics[width=0.75\textwidth]{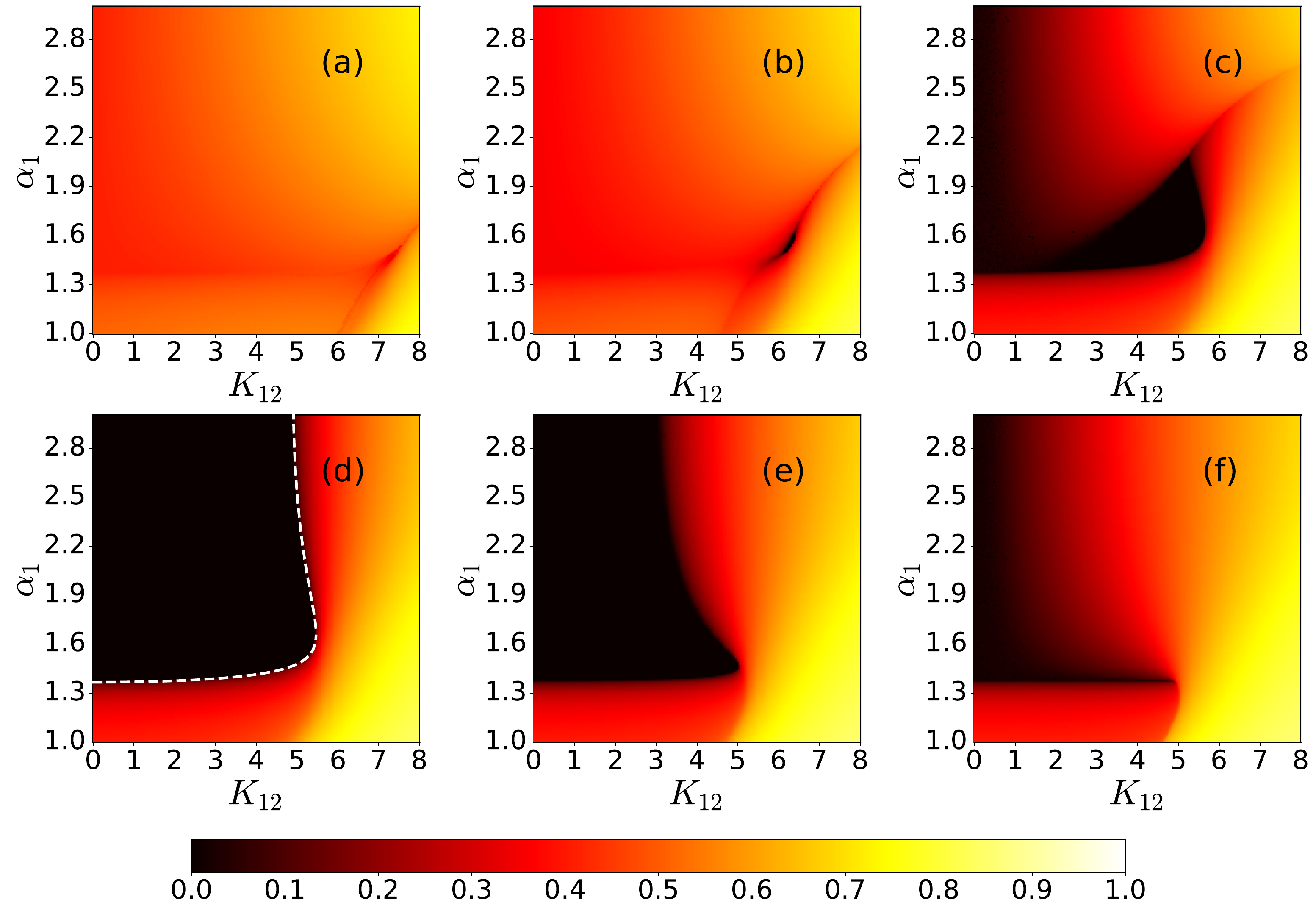}
	\caption{Heatmap in $\alpha_1 \times K_{12}$ parameter space showing the time averaged order parameter $\langle p_T \rangle$ for case 2 for different values of $J_2$:  (a) $J_2 = 6$; (b) $J_2 = 5$; (c)$J_2 = 2$; (d)$J_2 = 0$. The white dashed line corresponds to the boundary between the regions calculated via Eq. \eqref{eq::teorJ0}. The last two panels show results for case 2: (e) $K_2=1.2$; (f) $K_2=1.99$.  }
	\label{fig::ansatzK10J3VariaJ}
\end{figure}

The heatmaps show the first important feature of the dynamics for case 1, namely, the appearance of an island of fully asynchronous motion for large values of the inter-module coupling (around $K_{12}=5$ and $\alpha_{1}=1.5$), in which $\langle p_1 \rangle = \langle p_2 \rangle = 0$. This indicates a destructive interaction between the modules, where $\gum$ pulls $\gdo$ into fully disordered motion, even if it were strongly synchronized for $K_{12}=0$. Below this region both modules synchronize and above it we see chimera-like states, where $\gum$ is largely out of sync but $\gdo$ is synchronized. Regions of incoherent motion also occur for case 2, but for different parameters and only for small values of $K_2$ as we will show in Figure \ref{fig::ansatzK10J3VariaJ}.

\begin{figure}[htp]
\centering
\hfill
\begin{subfigure}[T]{0.43\textwidth}
\centering
\includegraphics[width=0.99\textwidth]{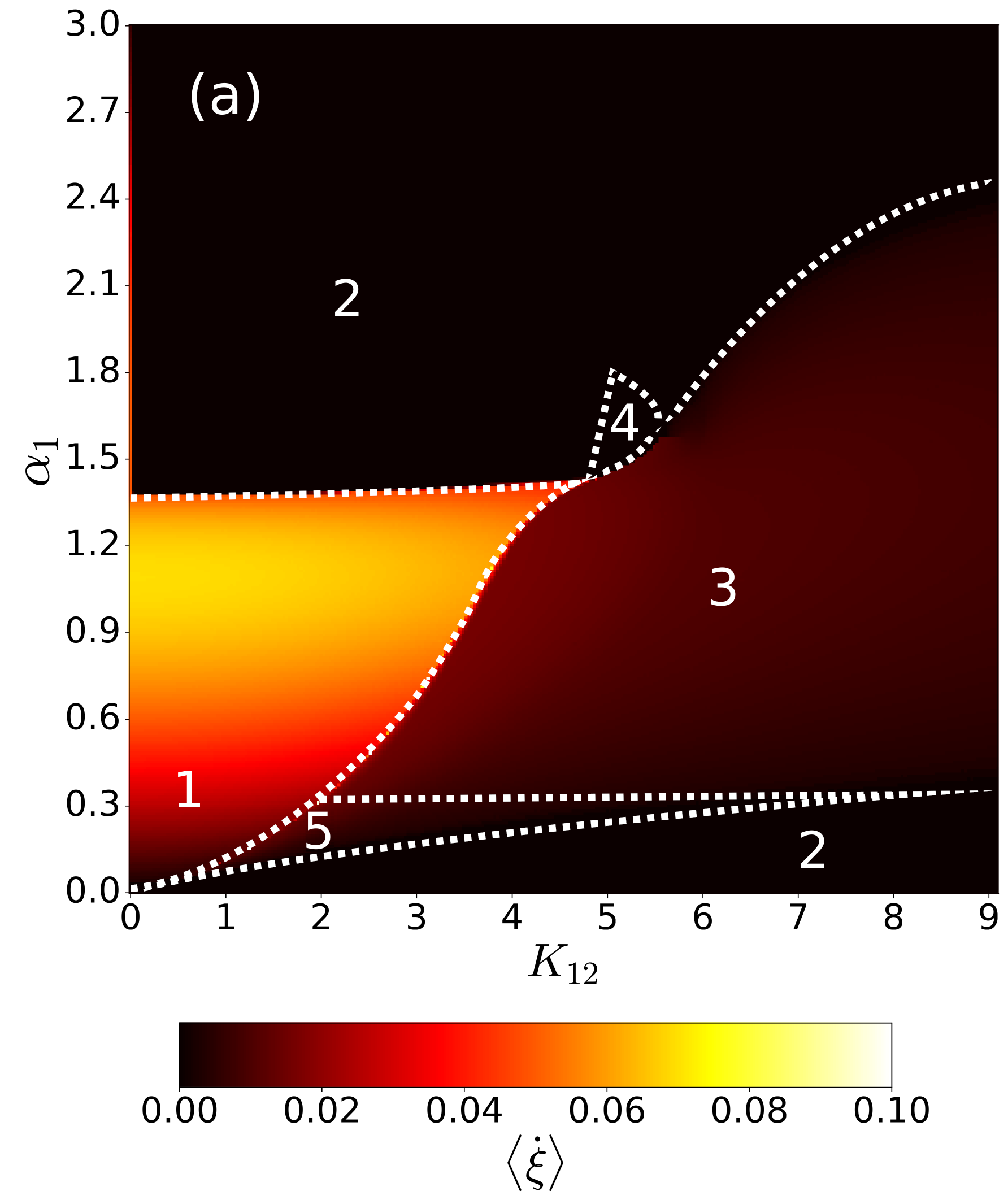}

\end{subfigure}
\hfill
\begin{subfigure}[T]{0.52\textwidth}
	\centering
	\includegraphics[width=\textwidth]{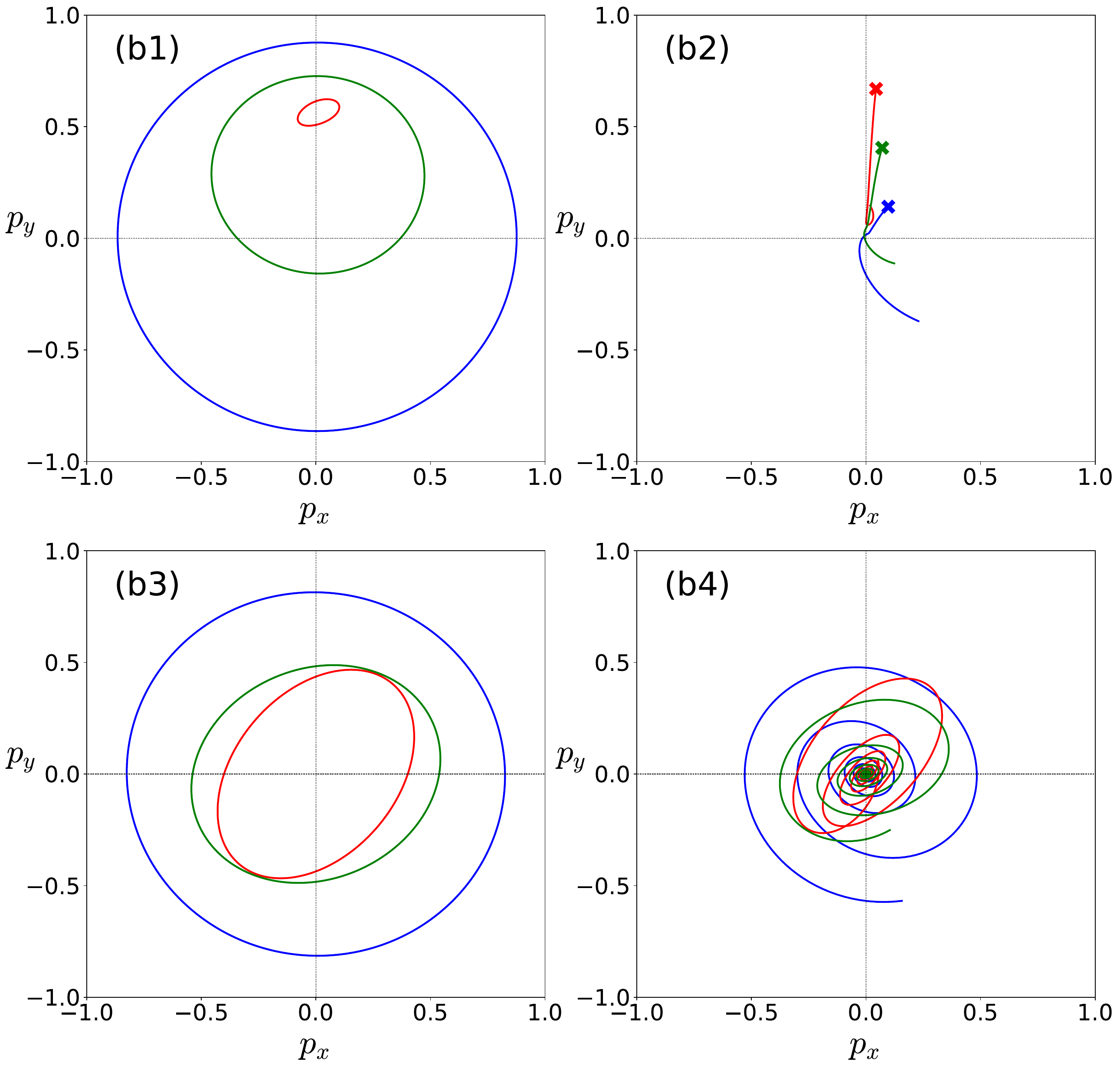}
\end{subfigure}
\caption{(a) Heatmap of the time averaged detuning $\langle \dot{\xi} \rangle$  in $\alpha_1 \times K_{12}$ parameter space for case 1. Panels (b1)-(b4) show examples of  trajectories $\vec{p}(t)$  illustrating the different dynamics that occur on the system. The parameter values that corresponds to each subpanel are indicated by numbers on (a), where dotted lines indicate approximate boundaries between the different types of behavior. In all cases, blue lines refers to $\gum$, red to $\gdo$ and green to $\gum + \gdo$. }
\label{fig::ansatzK10J3Freq}
\end{figure}

The time-averaged order parameter provides information about the system's synchronization intensity, but fails to identify the different synchronized patterns and the segregation/integration dynamics between the modules. Therefore, we also calculated the time averaged detuning between the modules  $\langle \dot{\xi} \rangle = \langle \dot{\psi_1} - \dot{\psi_2} \rangle$, to identify if the modules have distinct ($\langle \dot{\xi} \rangle \neq 0$) or similar ($\langle \dot{\xi} \rangle \approx 0$) dynamics. Figure \ref{fig::ansatzK10J3Freq}(a) shows $\langle \dot{\xi} \rangle$ as a heatmap in $\alpha_1 \times K_{12}$ plane and Figures\ref{fig::ansatzK10J3Freq} (b1) to (b4) show examples of trajectories $\vec{p}(t)$ for the system with $\mathcal{G}_2$ in the phase tuned state (case 1).
 
 Well defined partitions of the heatmap diagram in Figure \ref{fig::ansatzK10J3Freq}(a) can be observed (highlighted with guiding dotted lines):  the orange/red region 1, in which $\langle \dot{\xi} \rangle$ is maximum, indicates independence of the modules, i.e., aside from small perturbations, the motion of $\gum$ and $\gdo$ are equivalent to the uncoupled dynamics. Figure \ref{fig::ansatzK10J3Freq}(b1) illustrates the typical behavior of $\vec{p}$ in this region, in which $\vec{p}_1$ rotates uniformly, covering all quadrants while $\vec{p}_2$ suffers small perturbations but remains locked on the equilibrium point at $\pi/2$. The total behavior of the system looks like rotations with center displaced in the direction of  $\vec{p}_2$. The black region 2, in which $\langle \dot{\xi} \rangle = 0$, indicates a high level of integration between $\gum$ and $\gdo$, i.e., both modules with similar dynamics. By inspecting the trajectories of $\vec{p}$, such as in Figure \ref{fig::ansatzK10J3Freq}(b2), it can be observed that the dominating module is $\gdo$, since the whole system is now phase tuned. Although the example was picked from the top portion of $\alpha_1 \times K_{12}$ space, the same behavior was observed in the whole region 2. The brownish region 3 also presents small values of $\langle \dot{\xi} \rangle$, but not zero, indicating a weaker integration than in region 2. By analyzing the trajectories of $\vec{p}$ in Figure \ref{fig::ansatzK10J3Freq}(b3), it can be seen that that the dynamic is dominated by $\gum$ and both modules rotate with similar frequencies. The trajectories illustrated in Figure \ref{fig::ansatzK10J3Freq}(b4) refer to the asynchronous island, region 4, showing that $\vec{p}$ spirals to zero in the entire black region at the center of Figure \ref{fig::ansatzK10J3}. 
 
\begin{figure}[ht]
	\centering
	\includegraphics[width=0.8\textwidth]{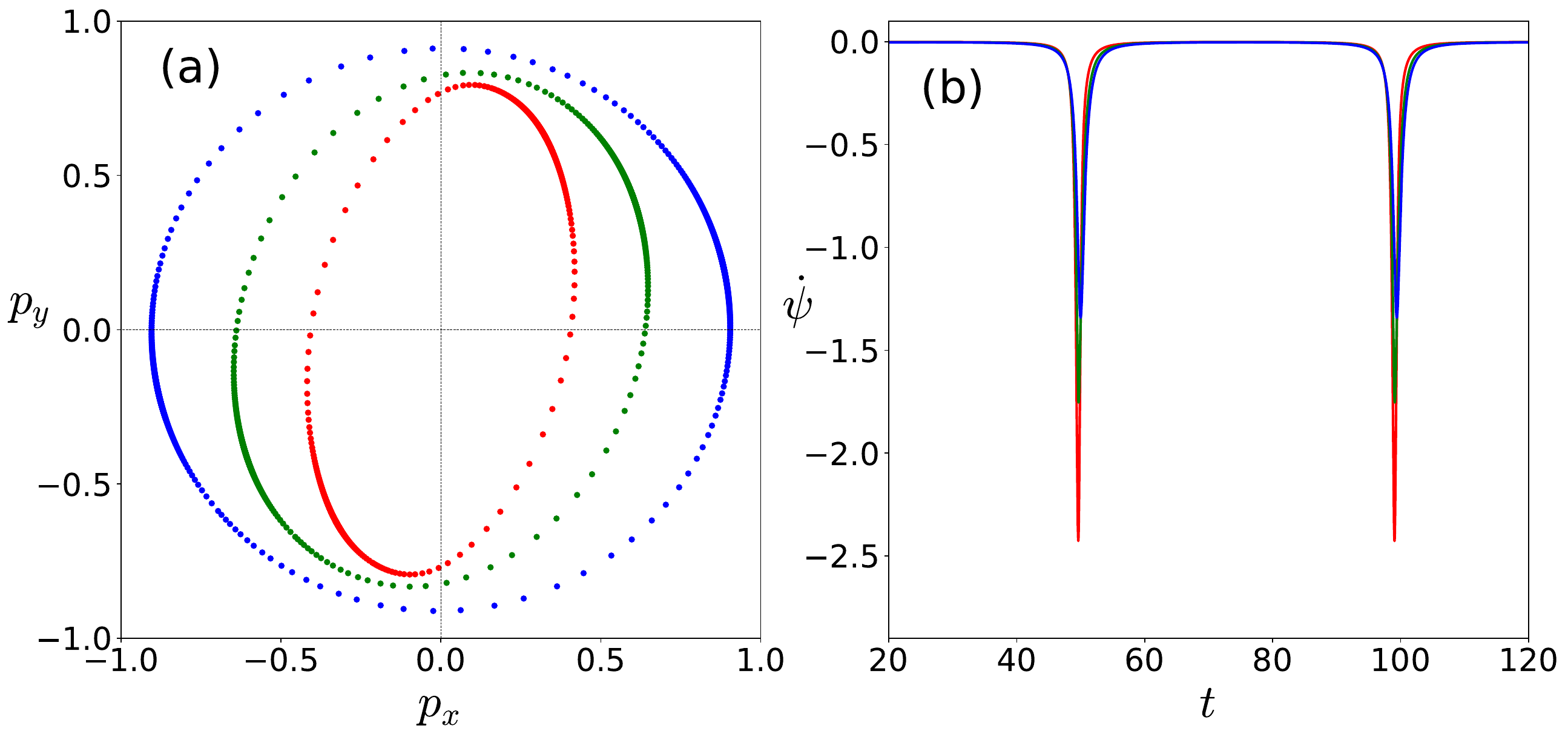}
	\caption{(a) Trajectories of the order parameter $\vec{p}$ in region $5$ of Figure \ref{fig::ansatzK10J3Freq}(a) and; (b) instantaneous frequency for the trajectories in (a). In all cases, blue lines refers to $\gum$, red to $\gdo$ and green to $\gum + \gdo$. The behavior resembles an excitatory neuron, with the angular velocities are nearly constant for long intervals ({\it charging}) with abrupt variation when $p_y \approx 0$ ({\it fire}). }
	\label{fig::fireP}
\end{figure}

The region marked with the number $5$, despite being relatively small, presents important behavior. Figure \ref{fig::fireP}(a) displays trajectories  of the order parameter $\vec{p}$ with points equally spaced in time, so that the density of points is proportional to the time the trajectories spend at each region. The density of points is clearly non-uniform, showing that oscillations of the order parameter are slow when $p_x$ is close to zero ($\psi \approx \pi/2$) and very fast  in the orthogonal direction, resembling the charge and fire dynamics of excitatory neurons. The spiking pattern of the instantaneous frequency $\langle \dot{\xi} \rangle$, Figure \ref{fig::fireP}(b) confirms this behavior. Although this type of dynamics can be modeled by simple equations for a single oscillator \cite{strogatz2018nonlinear}, here we demonstrate its occurrence for large groups of oscillators.

\begin{figure}[htp]
	\centering
	\hfill
	\begin{subfigure}[T]{0.33\textwidth}
		\centering
		\includegraphics[width=0.9\textwidth]{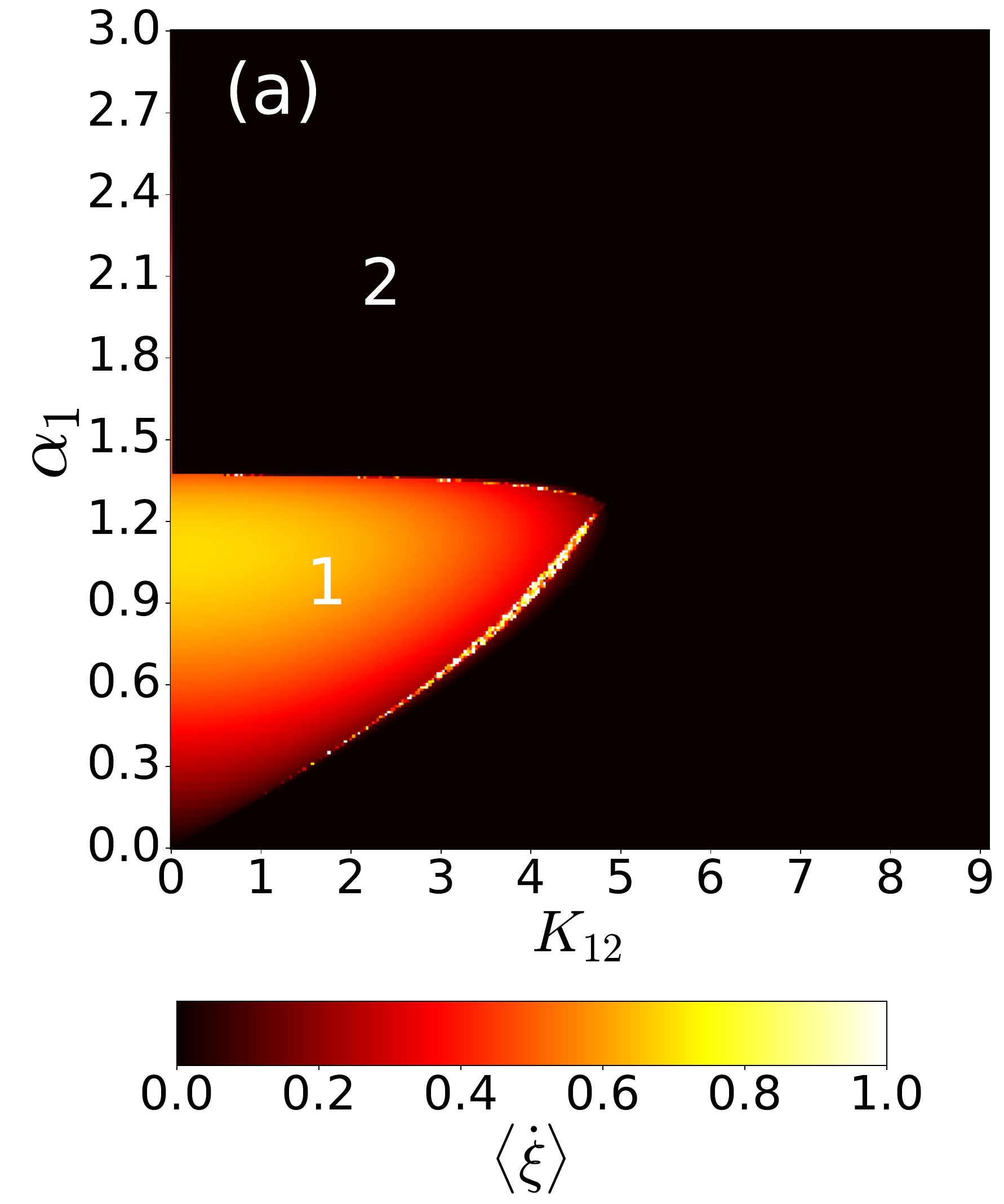}
		
	\end{subfigure}
	\hfill
	\begin{subfigure}[T]{0.66\textwidth}
		\centering
		\includegraphics[width=0.99\textwidth]{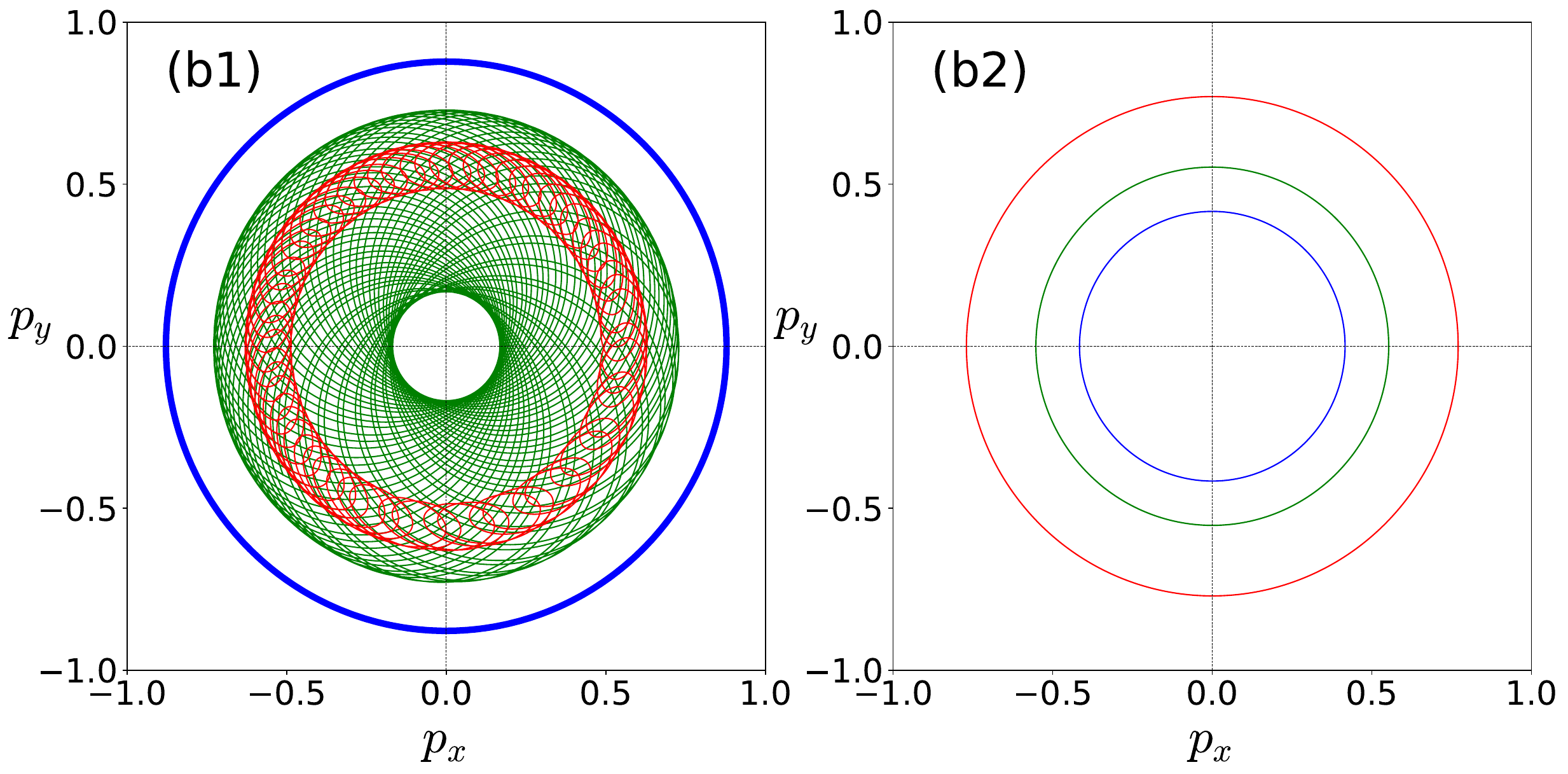}
	\end{subfigure}
	\caption{(a) Heatmap of the time averaged detuning $\langle \dot{\xi} \rangle$  in $\alpha_1 \times K_{12}$ parameter space for case 2. Panels (b1) and (b2) show examples of trajectories $\vec{p}(t)$  illustrating the different dynamics that occur in the system. The parameter values that corresponds to each subpanel are indicated by numbers in (a). In all cases, blue lines refers to $\gum$, red to $\gdo$ and green to $\gum + \gdo$.}
	\label{fig::ansatzK10K3Freq}
\end{figure}

In contrast, in Figure \ref{fig::ansatzK10K3Freq} we performed a similar analysis for case 2, and a much simpler diagram was found. In the orange region, in which $\langle \dot{\xi} \rangle > 0$, the modules are nearly independent, with dynamics resembling precessing oscillations. In the black region,  $\langle \dot{\xi} \rangle = 0$, the modules synchronize their oscillations. All non-trivial behaviors  found in case 1, such as the dessynchronization region and the excitation nonlinear oscillations only occur if $J_2 \neq 0$.

Next we investigate the impact of the intensity $J_2 $ in the asynchronous island found for case 1 (region 4 in Figure \ref{fig::ansatzK10J3Freq}). Heatmaps for the total order parameter $\langle p_T \rangle$ are shown in Figure \ref{fig::ansatzK10J3VariaJ} for different values of $J_2$. It can be seen that  the asynchronous region shrinks as $J_2$ increases, vanishing completely for $J_2 = 6$. The case of $J_2 = 0$, Figure \ref{fig::ansatzK10J3VariaJ}(d), is particularly interesting, as the dynamical equations become simple enough for analytical treatment. It corresponds to module $\gdo$ composed of free oscillators, driven only by their connections with $\gum$. To calculate the asynchronous  boundary we set $J_2 = 0$ in Eqs. \eqref{eq::ansatzP1} to \eqref{eq::ansatzPhi2} and combine equations for $\dot{\psi}_1$ and $\dot{\psi}_2$ into one for the phase difference $\dot{\xi} = \dot{\psi}_1 - \dot{\psi}_2$ to obtain
\begin{align*}
	\dot{p}_1 &= -\Delta p_1 + \frac{p_1}{2}(1-p_1^2) K_1 \operatorname{cos} \alpha_1
	+ \frac{p_2}{2}(1-p_1^2) K_{12} \operatorname{cos}\xi, \\
	\dot{p}_2 &= -\Delta p_2 + \frac{p_1}{2}(1-p_2^2) K_{12} \operatorname{cos}\xi, \\		
	\dot{\xi} &= - \frac{1+p_1^2}{2} K_1\operatorname{sin}\alpha_1 - \dfrac{K_{12} \sin{\xi}}{2} \left[ \frac{p_2}{p_1}(1+p_1^2) +\frac{p_1}{p_2}(1+p_2^2) \right] .
\end{align*}	

In equilibrium, when $p_1 \to 0$  and $p_2 \to 0$, we can approximate $(1\pm p_1^2) \approx 1$ and $(1 \pm p_2^2) \approx 1$. Defining the auxiliary ratio $g = p_2/p_1$, the equations can be written as
\begin{align}
	-2g\Delta &+ g K_1 \operatorname{cos} \alpha_1
	+ K_{12} \operatorname{cos}\xi   = 0,\\
	-2\Delta &+ gK_{12} \operatorname{cos}\xi  = 0,\\		
	- K_1\operatorname{sin}\alpha_1& - K_{12} \sin{\xi}\left[ g +\frac{1}{g} \right] = 0 .
\end{align}
Now, we can eliminate $g$ and $\cos \xi$ after some algebraic manipulation and write a single equation that involves variables $\cos \alpha_1$ and $K_{12}$ and parameters $K_1$ and $\Delta$:
\begin{equation}
	4\Delta^2 K^2_1 - 64\Delta^3 K_1\cos{\alpha_1} + 64\Delta^4 - K_{12}^2(K_1\cos{\alpha_1}-4\Delta)^2 - 2\Delta K^3_1 \cos{\alpha_1}+ 16\Delta^2 K^2_1 \cos^2{\alpha_1}= 0.
	\label{eq::teorJ0}
\end{equation}
Therefore, for a given set of parameters ($\Delta$;$K_{12}$;$K_1$), one can solve Eq. \eqref{eq::teorJ0} to find $\alpha_1$ that delimits the asynchronous region. The resulting curve is plotted  in Figure \ref{fig::ansatzK10J3VariaJ}-(d) as a dashed line and  perfectly delimits the asynchronous region.

We note that this calculation for case 1 and $J_2=0$ also holds for case 2 and $K_2=0$, as oscillators in module $\gdo$ are uncoupled (within the module) in both cases. Increasing $K_2$ from zero makes the asynchronous region shrink fast towards the lines $K_{12}=0$ and $\alpha_1 \approx 1.37$, disappearing for $K_2 \approx 2$, as shown in Figures. \ref{fig::ansatzK10J3VariaJ} (e) and (f) . This is very different from case 1, where an island of total decoherence survives for relatively large values of $K_{12}$.

\section{Simulations}
\label{sec::simulations}

Although the Ott-Antonsen ansatz enabled several analytical treatments and expanded the computational limit for numerical calculations \cite{Ott2008,childs2008,Skardal2020,Barioni2021,Buzanello2022}, the constraint of Lorentz distribution of natural frequencies imposed by the ansatz restricts its range of applications. Therefore, to further support our analysis, we performed direct simulations with oscillators using Gaussian distributions of natural frequencies.

\begin{figure}[ht]
	\centering
	\includegraphics[width=0.9\textwidth]{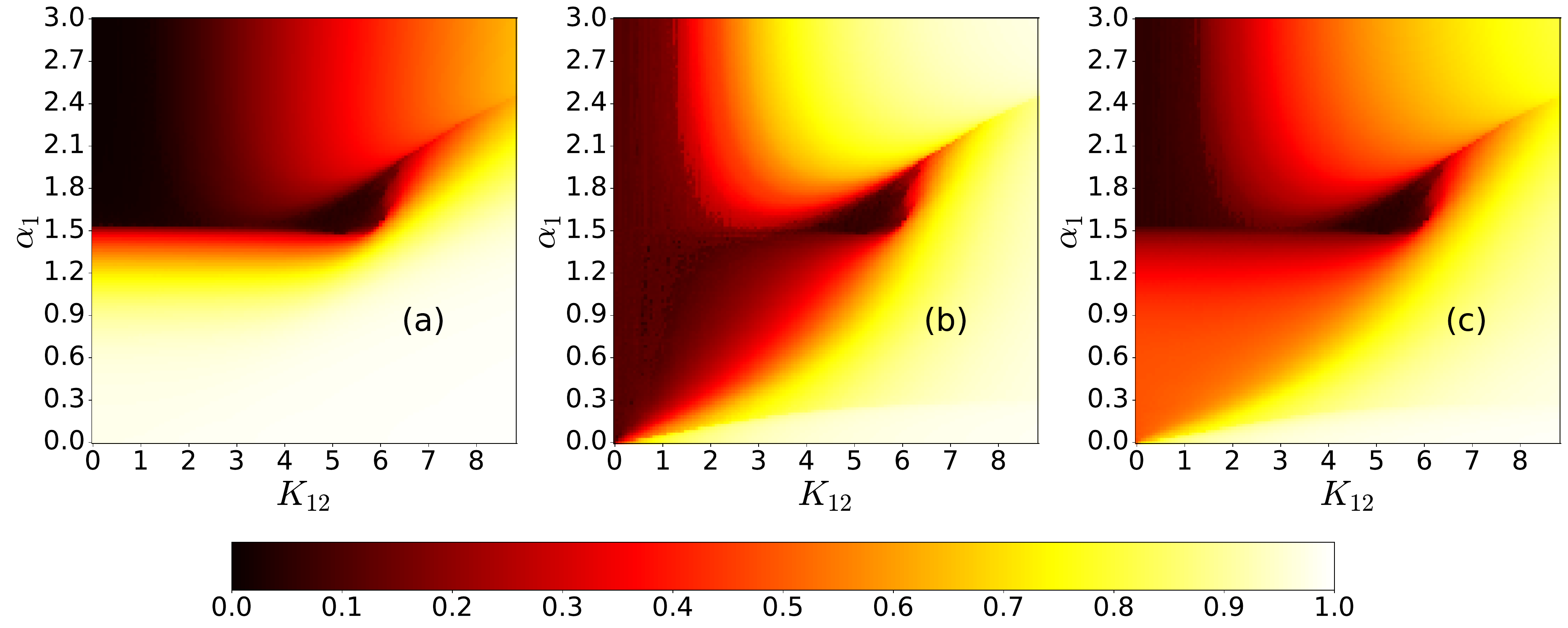}
	\caption{Time averaged order parameter $\langle p \rangle$ for a complete graph divided into two modules of $N=5000$ oscillators each with natural frequencies extracted from a Gaussian distribution with average $\omega_0 = 0$ and variance $\Delta = 1$. (a) Module 1 ; (b) Module 2 and (c) Both modules. }
	\label{fig::completeGraph}
\end{figure}

The first scenario mimics case 1 of the previous section, a complete graph with $N = 10000$ oscillators split into two groups with $\gum$ characterized by $K_1 = 10$, $J_1=0$ and $\alpha_1 \neq 0$ and $\gdo$ with $K_2=0$ and  $J_2 = 3$. Integrating Eq. \eqref{eq::dynamic2} for the oscillators, we calculated the time averaged order parameter $\langle p \rangle$ and constructed the heatmaps in $\alpha_1 \times K_{12}$ space, shown in Figure \ref{fig::completeGraph}. The general dynamical properties of the system as a whole are very similar to those found by the ansatz, including the excitable behavior displayed in Figure \ref{fig::fireP} and the asynchronous island, in the same approximate shape and position as in Figure \ref{fig::ansatzK10J3}. This indicates that these behaviors are somehow general and robust to other frequency distributions. 

Going one step closer to real systems, we relaxed the all-to-all interaction considered previously and performed simulations on synthetic modular networks. These graphs were built using the random partition algorithm \cite{Brandes2003}, in which we assign $N$ nodes to $\Omega$ partitions and try to connect every pair $(i,j)$ with probabilities $p_{in}$ or $p_{out}$ if nodes are within the same partition or not, respectively. For these simulations, we choose $N = 1000$, two partitions and average degree $\langle k \rangle = 10$. Therefore, the connection probabilities satisfy
\begin{equation}
	\dfrac{N(N-1)}{4} p_{in} + \dfrac{N^2}{4} p_{out} = \langle k \rangle.
\end{equation}

\begin{figure}[ht]
	\centering
	\includegraphics[width=0.9\textwidth]{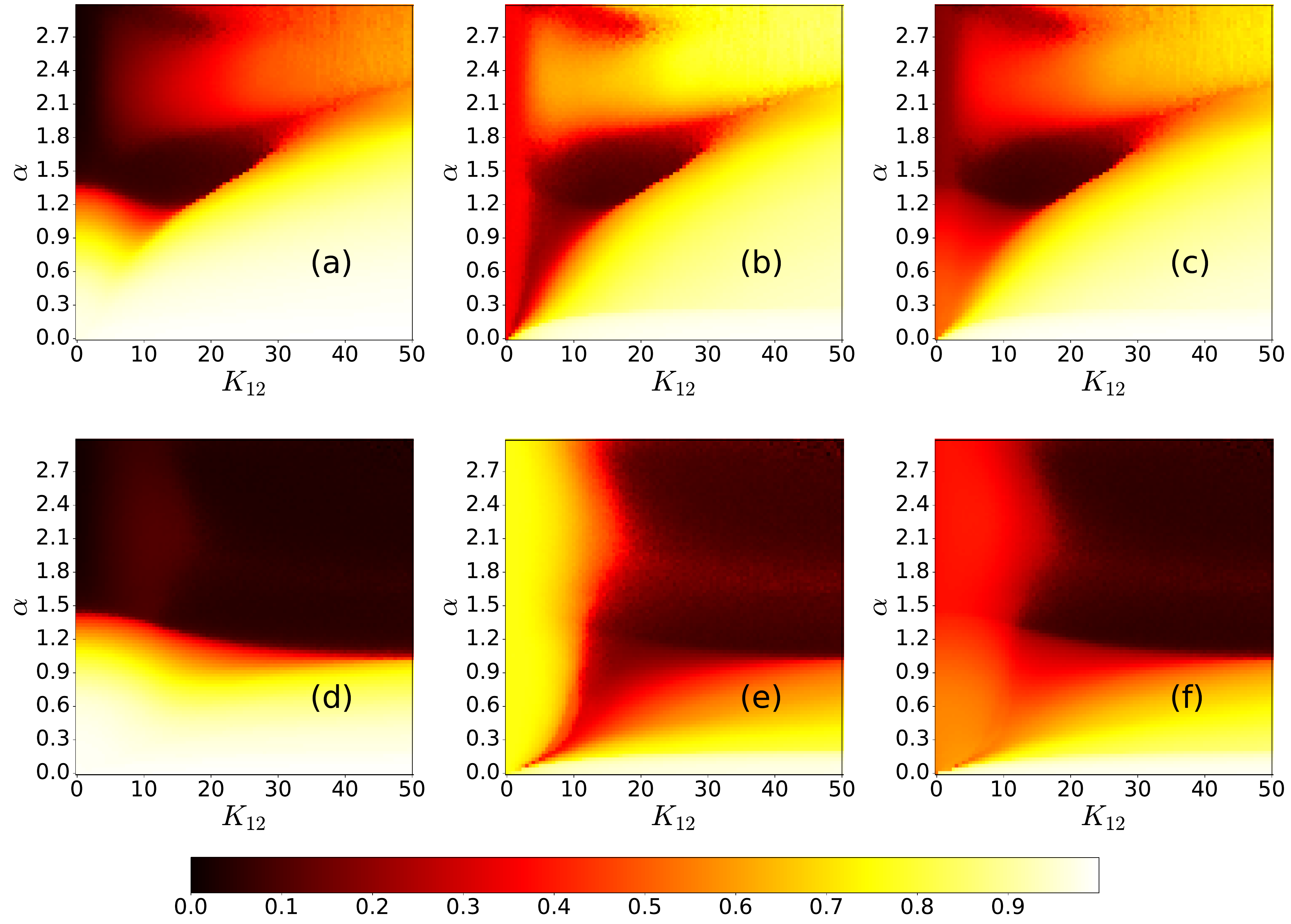}
    \caption{Time averaged order parameter $\langle p \rangle$ for a random partition graph divided into two modules summing $N=1000$ oscillators with natural frequencies extracted from a Gaussian distribution with average $\omega_0 = 0$ and variance $\Delta = 1$. $\kk{1}$ consists of $K_1 = 10$ and variable $\alpha_1$ while $\kk{2}$ consists only on $J_2 = 3$. (a) $\gum$ and $\lambda=3$ ; (b) $\gdo$ and $\lambda=3$ ; (c) $\gum + \gdo$ and $\lambda=3$ ; (d) $\gum$ and $\lambda=5$ ; (e) $\gdo$ and $\lambda=5$ ; (f) $\gum + \gdo$ and $\lambda=5$.}
	\label{fig::rndPart}
\end{figure}

We set $p_{in} = \lambda p_{out}$ and vary $\lambda$ in order to investigate the effects of connection density between modules in the equilibrium states. We set the same parameters as in case 1: $K_1 = 10$, variable $\alpha_1$ and $J_2 = 3$ and performed the numerical integrations, calculating $\langle p \rangle$ and constructing the heatmaps shown in Figure \ref{fig::rndPart}. For $\lambda = 3$, that corresponds to Figures \ref{fig::rndPart}-(a) to (c), the caracteristic asynchronous region presented in previous analysis can be seen clearly. It is important to notice that since the average number of connections between oscillators is small, the values of $K_{12}$ necessary to observe the different behaviors are much larger than in the all-to-all system. The shape of the heatmaps is qualitatively similar to that obtained by the ansatz, aside from the emergence of a second asynchronous region for $\alpha_1$ close to 3. Excitable trajectories similar to those in Figure \ref{fig::fireP} are once again found in the lower left corner of the diagram. Increasing $\lambda$ to $5$, Figures \ref{fig::rndPart}-(d)-(f), the dilution of edges between modules drastically changes the heatmaps, with a much larger disordered region. In addition, $\langle p \rangle$ becomes independent of intramodule coupling for $K_{12} < 10$. In this situation, the fraction of edges connecting oscillators from $\gum$ to module $\gdo$ is so small that modules can be considered independent. Further increasing $\lambda $ leads to similar heatmaps with the asynchronous region shifted to larger values of $K_{12}$ (not shown). We conjecture that for $\lambda$ large enough, the mutual influence of the modules will be insignificant to change their dynamics, no matter how large $K_{12}$ is.

\section{Discussion and Conclusions}
\label{sec::discussion}

One of the main features of the matrix coupled Kuramoto model is the existence of phase tuned solutions, where the phase of the order parameter converges to the direction of the dominant eigenvector of the coupling matrix. The matrix breaks the rotational symmetry and drives the order parameter to the direction of the dominant  eigenvector if its eigenvalues are real. In this paper we studied the synchronization dynamics of oscillators in modular networks where each module is governed by a different coupling matrix. 

We provided a detailed analysis for the case of two modules, where module $\gum$ followed the Kuramoto-Sakaguchi dynamics, with a rotating order parameter, and module $\gdo$ was in the phase tuned state. We identified five major regimes for the system's dynamics: a region in which both modules behave nearly independently; a region where $\gum$ dominates, so that both modules rotate at similar frequencies; a third region where $\gdo$ dominates and both modules tune their phases to a specific direction; and an asynchronous region where interaction between the modules is destructive. Finally we identified a fifth region where the system exhibits non-linear behavior typical of charge-and-fire of neurons, where the order parameter rotates slowly most of the time, with bursts of fast rotation. Interestingly, this behavior results from the collective dynamics of the whole network, that acts as a large excitable unit. If the dynamics of $\gdo$ is replaced by a simple Kuramoto dynamics, only the regime where $\gum$ dominates is possible. In this case a region of full asynchrony also develops, but only for small values of $K_2$ and $K_{12}$.
	
In order to investigate the robustness of our findings, we also performed simulations for oscillators with Gaussian distribution of natural frequencies, in which the Ott-Antonsen ansatz is no longer applicable. For complete graphs, in which interactions between oscillators are all-to-all, the same dynamics as those obtained by the ansatz were found. For random modular graphs, in which the average number of links between oscillators is small, the qualitatively behavior was confirmed for sufficiently connected modules. However, if the modules are weakly connected, they always behave independently, no matter how strong the intermodular coupling is.
	
A natural prospect of this study is to consider systems with more modules and an ensemble of coupling matrices, that may lead to even more interesting and richer dynamics. In addition, the use of real modular graphs, such as neuronal networks, as substrates for matrix coupled oscillators may give new insights into the synchronization phenomena. 
	
\begin{acknowledgments}
We thank Leonardo L. Bosnardo and Joao U.F. Lizarraga for helpful discussions. This work was partly supported by FAPESP [Grant 2023/03917-4 (G.S.C.)] and [Grant 2021/ 14335-0 (M.A.M.A.)] and by CNPq, Brazil, [Grant 301082/2019-7 (M.A.M.A.)].
\end{acknowledgments}






\end{document}